\documentclass[]{acm_sen_article}

\usepackage{lipsum}
\usepackage{flushend}

\usepackage{graphicx} 
\PassOptionsToPackage{hyphens}{url}
\usepackage{hyperref}
\usepackage{xspace}
\usepackage[numbers]{natbib}
\usepackage{balance}
\usepackage{xcolor}
\hypersetup{
  colorlinks=false,
  pdfborder={0 0 0}
}

\usepackage{fontawesome5}
\usepackage{framed}
\definecolor{formalshadelight}{RGB}{242,242,242}
\definecolor{formalshadedark}{RGB}{166,166,166}

\usepackage{csquotes}

\begin{document}

\title{ACM SIGSOFT SEN Empirical Software Engineering: Introducing Our New Regular Column}
\subtitle{ACM SIGSOFT SEN-ESE Column}

\numberofauthors{2} 
\author{
Justus Bogner\\
       \affaddr{Vrije Universiteit Amsterdam}\\
       \affaddr{The Netherlands}\\
       \email{j.bogner@vu.nl}
\and
Roberto Verdecchia\\
       \affaddr{University of Florence}\\
       \affaddr{Italy}\\
       \email{roberto.verdecchia@unifi.it}
}

\maketitle

\begin{abstract}
From its early foundations in the 1970s, empirical software engineering (ESE) has evolved into a mature research discipline that embraces a plethora of different topics, methodologies, and industrial practices.
Despite its remarkable progress, the ESE research field still needs to keep evolving, as new impediments, shortcomings, and technologies emerge.
Research reproducibility, limited external validity, subjectivity of reviews, and porting research results to industrial practices are just some examples of the drivers for improvements to ESE research.
Additionally, several facets of ESE research are not documented very explicitly, which makes it difficult for newcomers to pick them up.
With this new regular ACM SIGSOFT SEN column (SEN-ESE), we introduce a venue for discussing meta-aspects of ESE research, ranging from general topics such as the nature and best practices for replication packages to more nuanced themes such as statistical methods, interview transcription tools, and publishing interdisciplinary research.
Our aim for the column is to be a place where we can regularly spark conversations on ESE topics that might not often be touched upon or are left implicit. 
Contributions to this column will be grounded in expert interviews, focus groups, surveys, and position pieces, with the goal of encouraging reflection and improvement in how we conduct, communicate, teach, and ultimately improve ESE research.
Finally, we invite feedback from the ESE community on challenging, controversial, or underexplored topics, as well as suggestions for voices you would like to hear from.
While we cannot promise to act on every idea, we aim to shape this column around the community's interests and are grateful for all contributions.
\end{abstract}

\section{Introduction}
Empirical software engineering (ESE) emerged in response to the so-called \enquote{software crisis} of the 1960s, when growing system complexity revealed that theory alone was insufficient for effective software development~\cite{Gueheneuc2019}.
While the term \enquote{software engineering} was first popularized during the 1968 and 1969 NATO conferences~\cite{Naur1969}, it was only in the 1970s that researchers like Victor R. Basili at the University of Maryland began applying rigorous scientific methods to software engineering~\cite{Boehm2005}.
Basili’s pioneering work, which includes well-known frameworks such as Goal-Question-Metric~\cite{Basili94}, laid the foundations for empirical practices in the field, and guided software engineering on the path of becoming a more rigorous, evidence-based research field.
Over the years, many others have provided substantial methodological contributions to ESE research, e.g., Barbara Kitchenham for systematic literature reviews~\cite{Kitchenham2004}, Carolyn Seaman for qualitative studies~\cite{Seaman1999}, or Natalia Juristo~\cite{Juristo2001} and Claes Wohlin~\cite{wohlin_experimentation_2024} for controlled experiments, to only name a few (there are, of course, many others).

From the 80s and 90s into the 21st century, methodologies expanded from small-scale lab and student experiments to larger industrial case studies and controlled experimentation, tied together by efforts to establish reproducibility, replication, and stronger methodological robustness in software engineering research~\cite{felderer_evolution_2020}.
This was by no means an easy transformation, as many early SE empiricists received considerable pushback from SE researchers coming from a rationalist background~\cite{tichy_workings_2022}.
But over the past two decades, the ESE discipline has matured considerably: it now includes both quantitative and qualitative approaches, leverages data mining of large software repositories, employs systematic secondary studies and mixed-method designs, places strong emphasis on open science~\cite{Mendez2020}, and is supported by dedicated journals and conferences such as the Empirical Software Engineering journal (EMSE)\footnote{\url{https://link.springer.com/journal/10664}}, the International Symposium on Empirical Software Engineering and Measurement (ESEM)\footnote{\url{https://www.esem-conferences.org}}, and the International Conference on Evaluation and Assessment in Software Engineering (EASE)\footnote{\url{https://conf.researchr.org/series/ease}}.
Lastly, we also have high-quality, modern ESE books at our disposal, such as \citet{wohlin_experimentation_2024}, \citet{Felderer2020}, or \citet{Mendez2024}, an evolving collection of community-driven empirical standards~\cite{Ralph2020}, and have arguably become much better at the methodological education of the next generation of ESE researchers.

However, while we have made substantial progress as a field, there are still many things to improve.
Over the years, countless members of the community have pointed out various deficiencies in ESE research.
Some examples include:

\begin{itemize}
    \item the suboptimal reproducibility of ESE studies~\cite{Madeyski2017}, which has been pointed out several times in the last decades\footnote{\url{https://cacm.acm.org/blogcacm/the-rise-of-empirical-software-engineering-ii-what-we-are-still-missing/}},
    \item the difficulty of accumulating evidence on a phenomenon via replications due to the tendency of reviewers to expect substantial novelty and \enquote{surprising} findings~\cite{siegmund_views_2015, ernst_understanding_2021},
    \item the bad practice of drawing broad conclusions from low-quality or non-generalizable evidence~\cite{weyuker_empirical_2011},
    \item the inconsistent ESE education at universities that insufficiently prepares students for typical modern SE careers~\cite{Wilson2020},
    \item how the peer review processes of our flagship conferences tend to discourage innovative contributions on hard problems because they are too easy to reject\footnote{\url{https://sigsoft.medium.com/conferences-in-software-engineering-reflections-after-30-years-ca525d98c584}},
    \item how our unstructured peer review processes lead to subjective and unreliable paper evaluations~\cite{Arshad2021},
    \item and that the state of SE academia-industry bridges and SE science communication is worthy of improvement~\cite{Wilson2024, wyrich_beyond_2024}.
\end{itemize}
   
Despite our advances, empirical research in software engineering has by no means become easy, and demands much in terms of discipline, rigor, and expertise.
While we have more resources available than ever to support us, it is unavoidable that many things also remain unclear or are not sufficiently documented in an actionable way for junior researchers to pick up.
We recently noticed that there was little documented guidance for one of those things, namely how to write effective ESE papers.

\section{Notes on Writing ESE Papers}
After reading and writing the n-th ESE paper, its structure, content, and drafting process become almost second nature to many seasoned ESE researchers.
However, explaining all the nuances that ensure that an ESE study is well presented, especially to less experienced researchers such as Master and PhD students, remains difficult.
While the ESE community seems to have developed a reasonable shared understanding of how our ESE papers should roughly look like, we realized that this process is rarely documented and therefore hard to teach to newcomers.
To improve the situation, we started drafting a document detailing the common structure of ESE papers, the  content of their different sections, and general advice on how to write them effectively.
Our goal with releasing the final document, also published in ACM SIGSOFT SEN not long ago~\cite{verdecchia2025notes}, was not to set a new standard or lecture others on how ESE papers should be written.
Instead, it was to share how we are used to addressing a somewhat difficult and not very well-documented ESE topic, i.e., how to write effective ESE papers, in the hope that it would support early-career researchers and hopefully open a discourse on how the ESE community writes their papers.
We believe that there are many other ESE topics, e.g., some of the challenges mentioned in the introduction, that would benefit from a similar treatment.

\section{A Regular ESE Column: SEN-ESE}
To start making progress towards this goal, we have been invited to start a regular ACM SIGSOFT SEN column about empirical software engineering (SEN-ESE).
In the same spirit in which we released the \textit{Notes on Writing Effective ESE Papers}, we envision this column to be a place where we can regularly surface and spark conversations on ESE meta-topics that might not often be touched upon or are left implicit.
In the SEN-ESE column, we plan to cover aspects regarding all facets of ESE research, including but not limited to:

\begin{itemize}
    \item general ESE research, e.g., the nature of replication packages and best practices for creating them or publishing negative results within ESE venues,
    \item quantitative research, e.g., the role of hardware components in runtime metrics measurements or which usage of statistical methods is generally accepted in most ESE venues,
    \item qualitative research, e.g., transcription tools and guidelines for their usage or best practices for sampling interview participants,
    \item topics related to writing and presenting ESE research, e.g., how to write a great discussion section or how to make engaging presentations about ESE research,
    \item and potentially even community aspects, such as typical reviewing or editorial processes of ESE venues or conducting and publishing interdisciplinary ESE research.
\end{itemize}

Since we are far from being seasoned experts in most of these topics ourselves, we plan to ground our upcoming column articles in one of the following formats:

\begin{itemize}
    \item interview(s) or a focus group with experts on a topic,
    \item a questionnaire survey with the ESE community about a topic,
    \item or a position or opinion piece by us about a topic, with references to other ESE papers, and potentially with invited co-authors.
\end{itemize}

Overall, our goal with this regular column is not only to stimulate the ESE community to reason about how to effectively and systematically apply ESE research but also how we can improve our community processes and research methods, adapt them, and eventually evolve them to improve the ESE research field as a whole.
We hope that this column can ignite conversations that will ultimately lead to improving how ESE research is designed, executed, and reported.

\section{Call for Feedback and Participation}
While we outlined several concrete ideas above on what to cover in our column, we also want to explicitly request feedback from the ESE community, especially from more junior researchers, e.g., PhD students and postdocs.
What are ESE topics that you still find challenging, even despite available documented guidance, or that you think could benefit from more nuanced, opinionated insights from an expert in the field?
What are more controversial ESE topics that you would like to see discussed between experts of differing opinions?
What are important topics that are rarely openly discussed or published about in the ESE community and about which you are curious?
Are there well-known ESE community members whose experiences or advice on a certain topic you would be interested to hear?
Let us know, and we will consider covering your suggestions in our column!
Similarly, if you feel like you have something to say about one of our mentioned topics or a different important ESE topic based on your expertise and experience, please also let us know.
We will then consider whether to invite you for an interview or focus group.
While we cannot promise to directly act on all received feedback, it is definitely our intention to shape this column according to broader community interests.
We are therefore grateful for all feedback we receive, even if we likely cannot implement all of it.
Thank you in advance, and we hope you will enjoy the SEN-ESE column!
We are definitely looking forward to it!

\section{Acknowledgements}
We kindly thank Daniel Graziotin, University of Hohenheim, and Marvin Wyrich, Saarland University, for reviewing an earlier version of this article.

\bibliographystyle{plainnat}
\bibliography{biblio}

\end{document}